\documentclass{jetpl}
\twocolumn

\usepackage[cp1251]{inputenc}

\title{BUST Search for Muon Neutrinos from the Gravitational Wave Event GW170817}

\rtitle{BUST Search for Muon Neutrinos from the Gravitational Wave \ldots}

\sodtitle{BUST Search for Muon Neutrinos from the Gravitational Wave Event GW170817}

\author{V.\,B.\, Petkov$^{*, +}$\/\thanks{vpetkov@inr.ru}, M.\,M.\,~Boliev$^*$,  A.\,V.\,~Butkevich$^*$,   I.\,M.\,~Dzaparova$^{*, +}$, M.\,M.\,~Kochkarov$^*$, A.\,N.\,~Kurenya$^*$, A.\,S.\,~Lidvansky$^*$, Yu.\,F.\,~Novoseltsev$^*$, R.\,V.\,~Novoseltseva$^*$, P.\,S.\,~Striganov$^*$, A.\,F.\,~Yanin$^*$} 

\rauthor{V.\,B.\, Petkov, M.\,M.\, Boliev, A.\,V.\, Butkevich ...}

\sodauthor{Petkov, Boliev, Butkevich, ...}

\address{$^*$ Institute for Nuclear Research of RAS, 117312 Moscow, Russia\\
$^+$Institute of Astronomy of RAS, 119017 Moscow, Russia\\}

\abstract{
Using data of the Baksan Underground Scintillation Telescope (BUST) we have made a search for muon neutrinos and antineutrinos with energies above 1 GeV coinciding with the gravitational wave event GW170817 that was recorded on August 17, 2017 by the Advanced LIGO and Advanced Virgo observatories. This is a first detection of the new type of events occurring as a result of a merger of two neutron stars in a binary system. A short gamma-ray burst GRB170817A accompanying this event is an evidence of particle acceleration in the source whose precise position was determined by detection of the subsequent optical signal. No neutrino signals were found with the BUST in the interval  $\pm 500$ s around the moment of the gravitational wave event GW170817, as well as during the next 14 days. The upper limits on integral fluxes of muon neutrino and antineutrino from the source are derived.
}


\begin{document}

\maketitle

{\bf Introduction.} 
The gravitational wave event GW170817 \cite{bib:Abbott_1} detected by the Advanced LIGO and Advanced Virgo observatories was interpreted as a result of merging neutron stars in a binary system. Identification of this event with the gamma-ray burst GRB170817A that was detected by Fermi GBM in 1.7 s after the merger has confirmed the hypothesis about merging neutron stars and produced a direct proof of connection between merging neutron stars and short gamma-ray bursts. Subsequent detection of afterglow of the source in a wide range of wavelengths of electromagnetic radiation has confirmed the interpretation of this event as neutron star merging \cite{bib:Multimess_1}. 

High energy neutrinos from the GW170817 event were searched for by the neutrino telescopes ANTARES, IceCube, and Baikal-GVD, as well as by the air shower array Pierre Auger and Super-Kamiokande underground detector (see \cite{bib:HENu_1} - \cite{bib:SK_1}).  All these experiments used two time intervals when searching for neutrino events. The first one, $\pm 500$ s relative to the merger moment (the maximum possible interval between a gravitational wave and neutrinos from cosmic gamma ray bursts), was used in \cite{bib:Baret}, \cite{bib:Kimura} in order to search for prompt and prolonged gamma ray emission. The second interval, lasting 14 days after the merger, was used to search for high energy neutrinos from a long-living magnetar produced by merging neutron stars of a binary system \cite{bib:Gao}, \cite{bib:Fang}. 

None of the above listed experiments has found a signal from the GW1709817 gravitational event. The upper limits on the integral fluxes of neutrinos and on the total energy of neutrinos emitted by the source were obtained.

{\bf Baksan Underground Scintillation Telescope.} 
The BUST is located in Baksan Valley (North Caucasus, Russia) within an underground laboratory at an effective depth of 850 m of water equivalent. It is a multi-purpose instrument designed for a wide range of studies into physics of cosmic rays and elementary particles, and neutrino astrophysics \cite{bib:Alekseev_79}, \cite{bib:Alekseev_98}. The telescope of dimensions $17\times 17\times 11$ m$^3$ consists of 4 horizontal and 4 vertical planes with scintillation counters, the total number of which in the BUST being equal to 3184. The standard scintillation counter is an aluminum container with dimensions $0.7\times 0.7\times 0.3$ m$^3$ filled with organic liquid scintillator using white spirit as a solvent $C_nH_{2n+2}$ ($n\approx 9$). Each scintillation counter is viewed by a single PM tube FEU-49 with a photocathode diameter of 15 cm. The most probable energy release of muons in the counter is 50 MeV. Each counter has four output signals. The PM anode signal is used for fixing the time of a plane triggering and for measuring its energy release up to 2.5 GeV. The current output (anode signal coming through an integrating circuit) is used for adjustment and control of PM gains. Signals from the 12$^{th}$ dynodes feed the inputs of pulse shape discriminators (the so called pulse channel) with threshold amplitudes 8 and 10 MeV for the inner and outer planes, respectively. The signal from the 5$^{th}$ dynode comes to the input of a logarithmic converter, where it is transformed into a pulse whose duration is proportional to logarithm of signal amplitude. The logarithmic channel allows one to measure the energy release in each detector within the energy range 0.5 -- 600 GeV. 

The data acquisition system is triggered by actuation of the pulse channel of any BUST counter. The count rate of such trigger is 17 s$^{-1}$. When a trigger appears all data about a given event come to the on-line computer where pre-processing of events is performed in order to get information about the current state of recording devices. GPS signals with a synchronization accuracy of 0.2 ms are used in order to reference events with the universal time. 
 
{\bf Experiment.}
The BUST design allows one to identify trajectories of the muons crossing the telescope and to determine the muon arrival direction. The angular resolution of the instrument is $\approx 1.6^{\circ}$. It should be noted that the arrival direction of a muon produced in a neutrino reaction with matter strongly correlates with the neutrino arrival direction. The root mean square angle between a muon and its parent neutrino equals $\simeq 3.7^{\circ}$ for the energy spectrum of atmospheric neutrinos. Since the calculated spectra of neutrinos from astrophysical sources are harder than those of atmospheric neutrinos, one could expect a smaller angular distance between arrival directions of recorded muons and the direction to an astrophysical object.

When detecting muons from the lower hemisphere ($\theta > 90^{\circ}$) one can exclude the background from muons penetrating underground (since all known components of cosmic rays are absorbed at a depth of several kilometers of rock), if the flux of back-scattered muons from above is less than the neutrino effect at the telescope depth. For the BUST depth the muon background is totally excluded when zenith angles are $\theta > 100^{\circ}$. When an object is located in the upper hemisphere one can also search for muon neutrinos in the given time interval, provided that the muon background for a given direction is small, i.e., for directions with sufficiently large thickness of matter. 

Separation of arrival directions of muons between the upper and lower hemispheres is realized using the time-of-flight method. The time resolution of the telescope, when the relative time of flight between two scintillation planes is measured, equals 3.5 ns, and it is mainly determined by counter properties. For single muons from the upper hemisphere the reconstructed value of inverse velocity 1/$\beta$ ($\beta = v/c$) lies within the range 0.7 $\div$  1.3 for 95\% of events \cite{bib:Andreyev79}.  The same interval, but with the negative velocity sign is used to select muons from the lower hemisphere, generated in neutrino interactions with matter (rock) below the telescope. The threshold energy of muon neutrinos detected by the BUST is determined by energy losses of muons crossing the telescope, and it is equal to 1 GeV for the used selection criteria. 

{\bf Search for muon neutrinos from GW170817.}
Figure 1 presents the location of the GW170817 event in the local coordinate system (H, Az), where H is the elevation angle above horizon, and Az is the azimuth angle reckoned from the south direction. Also shown is the source trajectory during the first day after recording the gravitational wave event. During 9 hours per day the source is located above horizon, the minimum thickness of matter for these directions being equal to $\approx 10^6$ g/cm$^2$ \cite{bib:Gurentsov84}. This thickness of rock at the place of telescope location corresponds to the path length of muons with energy $\approx 10^5$ GeV \cite{bib:Mikheev82}. Muon neutrinos from GW170817 were searched for in a circle with a radius of $5.0^{\circ}$. In a time interval    $\pm 500$ s around the merging moment not a single event was found, in accordance with the expected number of background events from cosmic ray muons.   

During 14 days following after GW170817 3 muon events were detected, which is also in agreement with expected number of background events (2.6) produced by cosmic ray muons. All these events came from the upper hemisphere at elevation angles ranging from $6^{\circ}$ to $23^{\circ}$.

\begin{figure}
\centering
\includegraphics[width=8cm,clip]{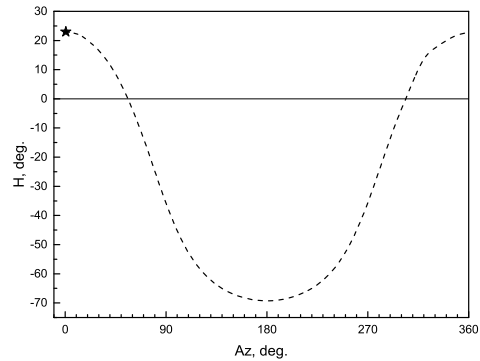}
\caption{Fig.1. The GW170817 event location (asterisk) and its trajectory in the local coordinate system (H, Az) during the first day after detection.}
\end{figure}    

{\bf Results.} 
From the fact of absence of neutrino signals from the source the upper limits (90\% confidence level) on integral fluxes of muon neutrinos and antineutrinos were derived as functions of their energy assuming monoenergetic spectrum (Fig. 2):
\begin{equation}
F(E_{\nu}) = \frac{N_{90}}{\epsilon S(E_{\nu})},
\label{Eq1}
\end{equation}
where $E_{min} = 1$ GeV, $E_{max} = 10^5$ GeV, $S(E_{\nu})$ is the effective area of detection of muon neutrino/antineutrino with energy $E_{\nu}$, $N_{90} = 2.3$, and   $\epsilon = 0.84$ is the portion of neutrino events from a point-like source within a circle of radius $5.0^{\circ}$. The limits are derived separately for muon neutrino and antineutrino, because interaction cross-sections are different for muon neutrinos and antineutrinos \cite{bib:Hayato09}, so that their effective areas of detection are different too.

\begin{figure}
\centering
\includegraphics[width=8cm,clip]{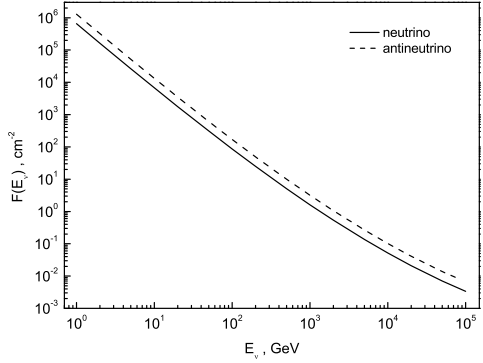}
\caption{Fig.2. The upper limits on integral fluxes of muon neutrinos and antineutrinos from GW170817 for the time interval $\pm 500$ s as functions of their energy (for mono-energetic spectrum).}  
\end{figure} 
 
Under assumption of the power-law spectrum with exponent $-2$, the upper limits on integral fluxes of muon neutrinos and antineutrinos from GW170817 are obtained for the energy range 1 GeV -- $10^5$ GeV as: 
\begin{equation}
F_{\nu} = \frac{N_{90}}{\epsilon \int\limits_{E_{min}}^{E_{max}}dE_{\nu}S(E_{\nu})I(E_{\nu})},
\label{Eq2}
\end{equation}
where $E_{min} = 1$  GeV, $E_{max} = 10^5$ GeV, and $I(E_{\nu})=E_{\nu}^{-2}$. 

In the energy range specified above the upper limits with 90\% confidence level are equal to 57.3 cm$^{-2}$ and 113.0 cm$^{-2}$ for muon neutrino and antineutrino, respectively. 

Figure 3 presents the upper limits on the energy fluxes in muon neutrino and antineutrino from GW170817 for the time interval $\pm 500$ s. The limits are calculated separately for every decade of energy and under assumption of a power-law spectrum with exponent $-2$:  
\begin{equation}
\Phi_{lim} = \frac{N_{90}\int\limits_{E_1}^{E_2}dE_{\nu}E_{\nu}I(E_{\nu})}{\epsilon \int\limits_{E_1}^{E_2}dE_{\nu}S(E_{\nu})I(E_{\nu})}.
\label{Eq3}
\end{equation}

\begin{figure}
\centering
\includegraphics[width=8cm,clip]{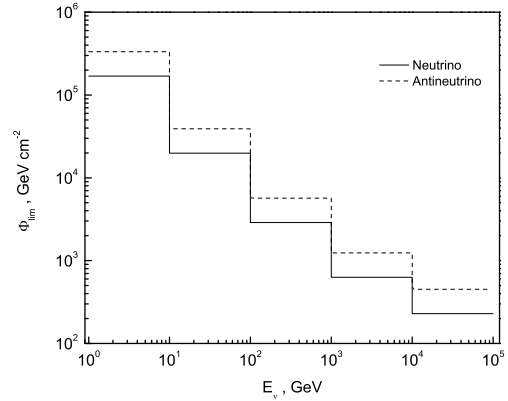}
\caption{Fig.3. The upper limits on energy fluxes from GW170817 for the time interval $\pm 500$ s.}
\end{figure}

\vspace{5mm} The work is performed with the Baksan Underground Telescope, Unique Scientific Facility of the Common Use Center Baksan Neutrino Observatory. It is supported by the Program of Fundamental Research ``Physics of hadrons, leptons, Higgs bosons and dark matter particles'' of the Presidium of Russian Academy of Sciences, and by the Russian Foundation for Basic Research (grant no. 17-52-80133).


\vspace{5mm}
\begin{thebibliography}{50}

\bibitem{bib:Abbott_1}
B.P. Abbott et al. (LIGO Scientific Collaboration, Virgo Collaboration), 
Phys. Rev. Lett. {\bf 119}, 161101, 2017; arXiv:1710.05832.

\bibitem{bib:Multimess_1}
B.P. Abbott et al. (LIGO Scientific Collaboration and Virgo Collaboration, Fermi GBM, INTEGRAL, IceCube Collaboration, AstroSat Cadmium Zinc
Telluride Imager Team, IPN Collaboration, The Insight-HXMT Collaboration, ANTARES Collaboration, The Swift Collaboration, AGILE Team, The 1M2H Team, The Dark Energy Camera GW-EM Collaboration and the DES Collaboration, The DLT40 Collaboration, GRAWITA: GRAvitational Wave Inaf TeAm, The Fermi Large Area Telescope Collaboration, ATCA: Australia Telescope Compact
Array, ASKAP: Australian SKA Pathfinder, Las Cumbres Observatory Group, OzGrav, DWF (Deeper, Wider, Faster Program), AST3, and CAASTRO Collaborations, The VINROUGE Collaboration, MASTER Collaboration, J-GEM, GROWTH, JAGWAR, Caltech-NRAO, TTU-NRAO, and NuSTAR Collaborations, Pan-STARRS, The MAXI Team, TZAC Consortium, KU Collaboration, Nordic
Optical Telescope, ePESSTO, GROND, Texas Tech University, SALT Group, TOROS: Transient Robotic Observatory of the South Collaboration, The BOOTES Collaboration, MWA: Murchison Widefield Array, The CALET Collaboration, IKI-GW Follow-up Collaboration, H.E.S.S. Collaboration, LOFAR Collaboration, LWA: Long Wavelength Array, HAWC Collaboration, The Pierre Auger
Collaboration, ALMA Collaboration, Euro VLBI Team, Pi of the Sky Collaboration, The Chandra Team at McGill University, DFN: Desert Fireball Network, ATLAS, High Time Resolution Universe Survey, RIMAS and RATIR, and SKA South Africa/MeerKAT),
Astrophys. J. L. {\bf 848}, L12, 2017.

\bibitem{bib:HENu_1}
A. Albert et al. (ANTARES Collaboration, IceCube Collaboration, The Pierre Auger Collaboration, and LIGO Scientific Collaboration and Virgo Collaboration),
Astrophys. J. L. {\bf 850}, L35, 2017; arXiv:1710.05839.

\bibitem{bib:Baikal}
A.D. Avrorin, A.V. Avrorin, V.M. Aynutdinov et al. (Baikal-GVD Collaboration),
Jetp Lett. {\bf 108}, 787, 2018; arXiv:1810.10966.

\bibitem{bib:SK_1}
K. Abe et al. (The Super-Kamiokande Collaboration), 
Astrophys. J. L. {\bf 857}, L4, 2018; arXiv:1802.04379.

\bibitem{bib:Baret}
B. Baret, I. Bartos, B. Bouhou et al., 
Astropart. Phys., {\bf 35}, 1, 2011; arXiv:1101.4669.

\bibitem{bib:Kimura}
S. Kimura, K. Murase, P. M\'esz\'aros, and K. Kiuchi,
Astrophys. J. L. {\bf 848}, L4, 2017.

\bibitem{bib:Gao}
H. Gao, B. Zhang, X. Wu and Zi. Dai,
Phys. Rev. D{\bf 88}, 043010, 2013; arXiv:1306.3006.

\bibitem{bib:Fang}
K. Fang and B.D. Metzger,
Astrophys. J. {\bf 849}, 153, 2017; arXiv:1707.04263.

\bibitem{bib:Alekseev_79}
E.N. Alekseev et al. (BUST Collaboration),  
Proceedings of 16th International Cosmic Ray Conference (Kyoto, Japan, August 6 - 18), {\bf 10}, 276, 1979.

\bibitem{bib:Alekseev_98}
E.N. Alekseev et al. (BUST Collaboration), 
Phys. Part. Nucl. {\bf 29}, 254, 1998. 

\bibitem{bib:Andreyev79}
Yu.M. Andreyev et al. (BUST Collaboration),
Proceedings of 16th International Cosmic Ray Conference (Kyoto, Japan, August 6 - 18), {\bf 10}, 184, 1979. 

\bibitem{bib:Gurentsov84}
V.I. Gurentsov, Preprint of Inst. for Nucl. Research, USSR Acad. Sci., no. 0379, Moscow, 1984.

\bibitem{bib:Mikheev82} 
S.P. Mikheev, Cand. Sci. (Phys.-Math.) Dissertation, Moscow: Inst. for Nucl. Research, USSR Acad. Sci., 1982. 

\bibitem{bib:Hayato09} 
Y. Hayato,
Acta Physica Polonica B {\bf 40}, 2477, 2009.




\end{thebibliography}
\end{document}